\documentclass[epj]{svjour2}
\usepackage{epsfig}
\def \sss {\hat{s}}
\def \tt {\hat{t}}
\def \uu {\hat{u}}
\def\sC{\mbox{\tiny C}}
\def\sD{\mbox{\tiny D}}
\def\sQ{\mbox{\tiny Q}}
\def\I{\mbox{\small i}}
\def\g{\mbox{g}}
\def\sms{\mbox{\scriptsize s}}
\def\sq{\mbox{\scriptsize q}}
\def\e{\mbox{e}}

\def\sc{\mbox{\scriptsize c}}
\def\sl{\mbox{\scriptsize l}}
\def\so{\mbox{\scriptsize o}}
\def\sr{\mbox{\scriptsize r}}

\begin{document}
\title{Hard exclusive baryon-antibaryon production\\ in two-photon
collisions}

\author{C.F. Berger\inst{1} \and W. Schweiger\inst{2}
}

 \institute{ C.N. Yang Institute for Theoretical Physics, SUNY Stony Brook,
Stony Brook, NY 11794-3840,
USA\and
 Institut f\"ur Theoretische Physik, Universit\"at Graz,
Universit\"atsplatz 5, A-8010 Graz,
 Austria }

\date{\today}

\abstract{We study baryon pair production in two-photon
collisions, $\gamma \gamma \rightarrow B \bar{B}$, within
perturbative quantum chromodynamics, treating baryons as
quark-diquark systems.  We extend previous work within the same
approach by treating constituent-mass effects systematically by
means of an expansion in the small parameter (mass/photon energy).
Our approach enables us to give a consistent description of the
cross sections for all octet baryon channels.  Adopting the model
parameters from foregoing work, we are able to reproduce the most
recent large-momentum-transfer data from LEP for the $p\bar{p}$,
$\Lambda\bar{\Lambda}$, and $\Sigma^0\bar{\Sigma}^0$ channels in a
quite satisfactory way.  We also briefly address the crossed
process for the proton channel,
$\gamma p \rightarrow \gamma p$.  \\
} 
\maketitle
\section{Introduction}
\label{sec:intro}

Theoretical analyses of exclusive reactions in quantum
chromodynamics (QCD), where intact hadrons appear in the initial
and final states, are of great importance for a better
understanding of the mechanism of confinement and of the
dynamics of hadronic bound states.  Although much progress has
been made in the theoretical understanding within frameworks
based on perturbation theory \cite{blrev,pire,stefanis}, there
still remain many open problems.  It is a matter of ongoing
discussion whether the currently experimentally accessible
energies are high enough such that the perturbative treatment
becomes applicable \cite{contro}.  As a possible way to model
non-perturbative effects which do not seem to be fully separated
from perturbatively calculable contributions at intermediately
large momentum transfers, the introduction of diquarks has been
proposed in \cite{anskrollpire}. In a series of papers this
effective model has been developed further and successfully
applied to a variety of exclusive reactions
\cite{gamgampp,Compton,fixed,time,Ja94,KSG96,KSPS97,fizb,CW,mass}.
In this work we continue the investigations within the diquark
model and consider exclusive two-photon reactions.

 The theoretical description of exclusive reactions with\-in
 perturbation theory is based on the ideas of Brodsky and Lepage
 \cite{blrev,blpaper}, and Efremov and Radyushkin \cite{efrad}.
 Within this so-called hard scattering picture (HSP), an exclusive
 reaction amplitude can be written as a convolution of
 process-dependent, perturbatively calculable, hard-scattering amplitudes
  with process-inde\-pen\-dent probability
 amplitudes for finding the pertinent valence Fock states in the
 scattering hadrons.  The latter are non-perturbative quantities, but
 their dependence on the momentum-trans\-fer $Q^2$ can be determined
 perturbatively.  Their extraction from experimental observations is
 challenged by the fact that they enter only integrated quantities,
 such as form factors.  However, their shape can be constrained with
 the help of QCD sum rules, lattice QCD and other non-per\-tur\-ba\-tive
 methods.  These studies seem to indicate that the distribution of
 longitudinal momentum fractions among the valence quarks in a nucleon
 is quite asymmetric for finite momentum transfers.  This can be
 interpreted as evidence for binding effects between two quarks in a
 nucleon and motivates the introduction of
 diquarks~\cite{stefanis,anselmino}.  Comprehensive reviews of the HSP
 are given, for example, in \cite{blrev,pire,stefanis}.

The HSP as presented above is exactly valid only for asymptotically
large momentum transfers, $Q^2 \rightarrow \infty$, where long- and
short-distance effects are completely incoherent.  However, as already
mentioned above, experimental observations seem to indicate that this
separation is not yet achieved at presently accessible momentum
transfers of a few GeV. Thus a perturbative calculation of the short
distance contributions may not be completely self-consistent.

Inspired by the aforementioned correlations observed in hadronic
wave functions, a quark-diquark model was developed in
\cite{anskrollpire} to parameterize possible non-perturbative
effects within a perturbative framework.  Within this model which
is based on the HSP, baryons are treated as quark-diquark systems.
The composite nature of the diquarks is taken into account by
diquark form factors which are parameterized such that
asymptotically the scaling behavior of the pure quark HSP emerges.
The possibility of the reformulation of the pure quark HSP in
terms of quark and diquark degrees of freedom has been
demonstrated in \cite{bernd,marc}.  In earlier studies of
two-photon annihilation into baryons, $\gamma \gamma \rightarrow B
\bar{B}$ \cite{gamgampp}, and of Compton scattering off baryons,
$\gamma B \rightarrow \gamma B$ \cite{Compton,KSG96}, all quark
masses have been neglected, while masses for diquarks were
introduced as additional parameters.  In recent studies within the
diquark model \cite{fizb,CW,mass} a different strategy was adopted
to treat mass effects more consistently without introducing new
mass parameters for the hadronic constituents. In the following we
will reconsider the two-photon reactions with this improved
treatment of mass effects.

We start by introducing the necessary ingredients of the diquark
model for the reaction $\gamma \gamma \rightarrow B \bar{B}$.
While we try to keep the present discussion as self-contained as
possible, we omit certain details of the model which can be found,
for example, in \cite{CW}.  We explain our choice of quark and
diquark distribution amplitudes (DAs), and our treatment of
constituent masses in Sections \ref{sec:amp.qD} and
\ref{sec:mass}, respectively.  Having collected all ingredients of
the model, we list our analytical results for the hard scattering
amplitudes for the two-photon annihilation process in Section
\ref{sec:ampres}. In Section \ref{sec:results} we present the
numerical results for this reaction, compare to existing data for
the proton-antiproton, $\Lambda \bar{\Lambda}$, and $\Sigma^0
\bar{\Sigma}^0$ channels, and give predictions for other final
states. We then go on and briefly comment on the crossed reaction,
specifically Compton scattering off protons. Section
\ref{sec:final} ends the discussion with a summary and some
concluding remarks.

\section{The $\gamma \gamma \rightarrow B \bar{B}$ amplitude}
\label{sec:amp}

As stated in the introduction, within the HSP an exclusive scattering
amplitude can be written as a convolution integral of a
hard-scattering amplitude with distribution amplitudes, describing the
longitudinal momentum distribution of valence quarks in the
participating hadrons.  In the diquark model a baryon is considered as
consisting of a quark and a diquark.  The diquark is treated as a
quasi-elementary constituent which may survive medium hard collisions.

Within this framework, we obtain the following convolution integral
for the $\gamma \gamma \rightarrow B \bar{B}$ amplitude:
\begin{eqnarray}
\mathcal{M}_{\{\lambda\} }\left(\sss, \tt\right) & = &
\int\limits_0^1 d x_1 \int\limits_0^1 d y_1 \Psi_B^\dagger
\left(x_1\right)
\Psi_{\bar{B}}^\dagger \left(y_1\right) \nonumber \\
& \times & \hat{T}_{\{\lambda\} } \left(x_1, y_1; \sss,
\tt\right), \label{HSP}
\end{eqnarray}
where we have suppressed Lorentz and color indices in the
convolution integral.  Here $\hat{T}$ is the process-dependent
hard-scattering amplitude for producing a quark-an\-ti\-quark and
a diquark-antidiquark pair in a two-photon collision.  In the
notation used above, the quark and the antiquark carry momentum
fractions $x_1$ in the ba\-ryon, and $y_1$ in the antibaryon,
while the diquark and antidiquark carry momentum fractions $x_2 =
1-x_1$ and $y_2 = 1-y_1$, respectively ($0 \leq x_i,\, y_i \leq
1$).  $\hat{T}$ consists of all possible tree diagrams
contributing to the elementary scattering process $\gamma \gamma
\rightarrow q D \bar{q} \bar{D}$, where the momenta of all
constituents (quarks $q$ and diquarks $D$) are collinear to those
of their parent hadrons.  The distribution amplitudes $\Psi_{B}$
are process-independent probability amplitudes for finding the
pertinent valence Fock states in the baryon $B$ with the
constituents carrying the longitudinal momentum fractions
$x_i,\,y_i$ of the parent baryon and being collinear up to the
(factorization) scale $\tilde{p}_\perp$. In (\ref{HSP}) we have
already neglected the (logarithmic) $\tilde{p}_\perp$ dependence,
since this dependence is only of minor importance in the
restricted kinematic range of intermediately large momentum
transfer we are considering in this work. For convenience, we use
massless Mandelstam variables $\sss,\, \tt$, and $\uu$, rather
than the usual (massive) ones, $s,\, t$, and $u$. The subscript
${\{\lambda\} }$ denotes all possible configurations of photon and
baryon helicities.

For the process $\gamma \gamma \rightarrow B \bar{B}$, there are six
independent complex helicity amplitudes
$\mathcal{M}_{\lambda_{\bar{B}},\,\lambda_{B};\,\lambda_1,\,\lambda_2}$,
where the $\lambda_i,\,i = 1,2$ label the helicities of the incoming
photons, and $\lambda_{B},\,\lambda_{\bar{B}}$ are the helicities of
the baryon and antibaryon, respectively. Following the conventions in
\cite{rosti76}, we express our observables in terms of the following
amplitudes:
\begin{eqnarray}
\overline{\phi}_1 & = &
\mathcal{M}_{-\frac{1}{2},\,\frac{1}{2};\,1,\,-1},\nonumber \\
\overline{\phi}_2 & = &
\mathcal{M}_{-\frac{1}{2},\,-\frac{1}{2};\,1,\,1}, \nonumber \\
\overline{\phi}_3 & = &
\mathcal{M}_{-\frac{1}{2},\,\frac{1}{2};\,1,\,1}, \nonumber \\
\overline{\phi}_4 & = &
\mathcal{M}_{-\frac{1}{2},\,-\frac{1}{2};\,1,\,-1}, \nonumber \\
\overline{\phi}_5 & = &
\mathcal{M}_{\frac{1}{2},\,-\frac{1}{2};\,1,\,-1}, \nonumber \\
\overline{\phi}_6 & = &
\mathcal{M}_{\frac{1}{2},\,\frac{1}{2};\,1,\,1}. \label{cmsampscr}
\end{eqnarray}
The remaining helicity configurations are related to these via parity
and time reversal invariance. We normalize these helicity amplitudes
such that the differential cross section for two-photon annihilation
into a baryon-antibaryon pair is given by
\begin{eqnarray}
\frac{d \sigma }{d t } & = & \frac{ 1}{32 \pi s^2 } \left[ \left|
\overline{\phi}_1 \right|^2 + \left| \overline{\phi}_2 \right|^2 + 2
\left|
\overline{\phi}_3 \right|^2 \right. \nonumber \\
& & \quad  \quad +  \left.
2 \left| \overline{\phi}_4 \right|^2 + \left| \overline{\phi}_5
\right|^2 + \left|
\overline{\phi}_6 \right|^2
 \right].
\label{dsigmacross}
\end{eqnarray}

Within the HSP constituent masses are usually neglected.  We
relate constituent masses to the pertinent baryon mass when
calculating the hard-scattering amplitude.  By a subsequent
expansion in powers of $(m_B/\sqrt{\sss})$, where $m_B$ denotes
the baryon mass, we obtain the leading mass-correction terms.  We
will elaborate more on our treatment of mass effects in Section
\ref{sec:mass}.

\subsection{Baryons as quark-diquark systems}
\label{sec:amp.qD}

In the ground state diquarks have positive parity and either spin 0
(scalar diquarks $S$) or spin 1 (vector diquarks $V$).
Because vector diquarks are mainly responsible for spin flips they are
essential to describe spin effects.

Neglecting transverse momentum, we can write the baryon wave
function in a covariant way so that only baryonic quantities
(momentum $p_B$, helicity $\lambda$, baryon mass $m_{B}$) appear.
For the lowest-lying baryon octet, assuming zero relative orbital
angular momentum between quark and diquark, the wave function
(already integrated over transverse momenta) can be written as
\begin{eqnarray}
\label{su6wf}
\Psi_{B}(p_{B}, x, \lambda) & = & f_{S}^{B} \Psi_{S}^{B}
(x) \chi_{S}^{B} \,u(p_{B},\lambda) \label{covwav} \\
& + & f_{V}^{B} \Psi_{V}^{B}(x) \chi_{V}^{B} \frac{1}
{ \sqrt{3}}\left( \gamma^\alpha
+ \frac{p_{B}^\alpha }{m_{B} } \right) \gamma_5\,u(p_{B},\lambda).
\nonumber
\end{eqnarray}
Here the $\chi_{D}^{B}$ are SU(3) quark-diquark flavor wave
functions, and the two functions $\Psi_{S}$ and $\Psi_{V}$, for
scalar and vector diquarks, respectively, represent the
nonperturbative probability amplitudes for finding these
constituents in the baryon.  The notation in Eq.  (\ref{covwav})
is slightly sloppy, the open index in the vector part of the wave
function corresponds to the Lorentz index of the V-diquark
polarization vector and is contracted appropriately in the
convolution integral, Eq. (\ref{HSP}).  The constants $f_{D}^B$
may be interpreted as baryon decay constants.  The wave function
for antibaryons is given by
\begin{eqnarray}
\label{su6wfa}
\Psi_{\bar{B}}(p_{\bar{B}}, x, \lambda) & = & f_{S}^{B} \Psi_{S}^{B}
(x) \chi_{S}^{\bar{B}} \,v(p_{\bar{B}},\lambda) \label{covwavout} \\
& + & f_{V}^{B} \Psi_{V}^{B}(x) \chi_{V}^{\bar{B}} \frac{ 1} {
\sqrt{3}}\left( \gamma^\alpha - \frac{p_{\bar{B}}^\alpha }{m_{\bar{B}}
} \right) \gamma_5\,v(p_{\bar{B}},\lambda), \nonumber
\end{eqnarray}
where we have used charge \,conjugation \,$v(p,\lambda) = i \gamma^2$
$u^*(p,\lambda)$.

In the following we will use a quark-diquark distribution amplitude
of the form \cite{huang}
\begin{eqnarray}
\Psi_{D}^{B}(x) & = & N^{B} x (1-x)^3 (1 + c_1 x + c_2 x^2)
\nonumber \\
& & \qquad \qquad \quad \;
\exp \left\{-b^2 \left( \frac{ m_q^2}{x } + \frac{m_{D}^2 }{1-x
}\right)\right\},
\label{huangv}
\end{eqnarray}
where $c_i = 0$ for scalar diquarks.  This DA for octet baryons
$B$ has been successfully used in previous applications of the
diquark-model \cite{fixed,time,Ja94,KSG96,KSPS97,fizb,CW,mass}. It
is an adaptation of a meson DA, obtained by transforming a
harmonic oscillator wave function to the light cone.  Therefore,
the masses appearing in the exponent are constituent masses.  The
oscillator parameter $b$ is fixed by the requirement that the mean
intrinsic transverse momentum of quarks inside the baryon $B$ is
$\sqrt{\left< k_T^2 \right>} \approx$ 600 MeV. This value was
found experimentally by the EMC collaboration in semi-inclusive
deep inelastic $\mu p$ scattering \cite{EMC}.  Theoretical
considerations also indicate a value of this magnitude
\cite{sterman}.  Furthermore, the exponent suppresses
contributions from the endpoint regions $x \rightarrow 0, 1$ in
the convolution integral (\ref{HSP}).  Endpoint-damping
Sudakov-type exponents generally arise when resumming corrections
from soft gluon radiation \cite{blpaper,sudakov}.  The exponent in
(\ref{huangv}) could be interpreted as simulating this Sudakov
suppression effect in the endpoint region.  The normalization
constant $N^{B}$ in Eq. (\ref{huangv}) is fixed by the requirement
that
\begin{equation}
\int\limits_0^1 \Psi_{D}(x) d x = 1. \label{normalda}
\end{equation}

The SU(3) quark-diquark flavor wave functions $\chi_{S}^{D}$ for
the lowest-lying baryon octet are listed in Table \ref{flav}.
\begin{table}
\caption{SU(3) quark-diquark flavor wave functions for the lowest
lying baryon octet}
\label{flav}
\[ \begin{array}{l@{\hspace*{2mm}}l}
\hline\noalign{\smallskip}
\chi_{S}^p = u S_{[u,d]} & \chi_{V}^p = \frac{\scriptstyle 1}
{\scriptstyle \sqrt{3}}\left[ u V_{\{u,d\}} - \sqrt{2} d V_{\{u,u\}}
\right] \\
\chi_{S}^n = d S_{[u,d]} &  \chi_{V}^n = -\frac{\scriptstyle 1}
{\scriptstyle \sqrt{3}}\left[ d V_{\{u,d\}} - \sqrt{2} u V_{\{d,d\}}
\right] \\
\chi_{S}^{\Sigma^{+}} \!\!\! = - u S_{[u,s]} &  \chi_{V}^{\Sigma^{+}}
=
\frac{\scriptstyle 1}
{\scriptstyle \sqrt{3}}\left[ u V_{\{u,s\}} - \sqrt{2} s V_{\{u,u\}}
\right] \\
\chi_{S}^{\Sigma^{0}} \!\!\!\! = \!\! \frac{\scriptstyle
1}{\scriptstyle \sqrt{2}}
\left[ d S_{[u,s]} + u S_{[d,s]} \right]  &  \chi_{V}^{\Sigma^{0}} =
\frac{\scriptstyle 1}{\scriptstyle \sqrt{6}} \left[ 2 s V_{\{u,d\}} -
d V_{\{u,s\}} \right. \\
& \qquad \quad \left. - u V_{\{d,s\}} \right] \\
\chi_{S}^{\Sigma^{-}} \!\!\! = d S_{[d,s]} & \chi_{V}^{\Sigma^{-}} =
- \frac{\scriptstyle 1}
{\scriptstyle \sqrt{3}}\left[ d V_{\{d,s\}} - \sqrt{2} s V_{\{d,d\}}
\right] \\
\chi_{S}^{\Lambda} = \frac{\scriptstyle 1}{\scriptstyle \sqrt{6}}
\left[ u S_{[d,s]} - d S_{[u,s]} \right. &
\chi_{V}^{\Lambda} \! = \! \frac{\scriptstyle 1}
{\scriptstyle \sqrt{2}}\left[ u V_{\{d,s\}} - d V_{\{u,s\}} \right] \\
\qquad \qquad \left. \, - 2 s S_{[u,d]} \right] & \\
\chi_{S}^{\Xi^{0}} \!\!\! = s S_{[u,s]} & \chi_{V}^{\Xi^{0}} =
-\frac{\scriptstyle 1}
{\scriptstyle \sqrt{3}}\left[ s V_{\{u,s\}} - \sqrt{2} u V_{\{s,s\}}
\right] \\
\chi_{S}^{\Xi^{-}} \!\!\! = s S_{[d,s]} & \chi_{V}^{\Xi^{-}} =
-\frac{\scriptstyle 1}
{\scriptstyle \sqrt{3}}\left[ s V_{\{d,s\}} - \sqrt{2} d V_{\{s,s\}}
\right] \\
\noalign{\smallskip}\hline
\end{array} \]
\end{table}
In (\ref{su6wf}) (and (\ref{su6wfa})) we have omitted the color part of
the quark-diquark wave function, which is given by
\begin{equation}
\Psi_{\sq D}^{\sc \so \sl \so \sr} = \frac{1}{\sqrt{3}}
\sum\limits_{a = 1}^3 \delta_{a \bar{a}},
\end{equation}
since diquarks are in a color antitriplet state because bary\-ons are
color singlets.

The hard-scattering amplitudes are calculated perturbatively with
point-like constituents.  For sake of completeness, we list the
Feynman rules within the diquark model in the Appendix.  Vector
diquarks are allowed to possess an anomalous (chromo)magnetic moment
$\kappa_V$, corresponding to the most general form of the coupling of
a spin-1 gauge boson to a spin-1 particle.  The composite nature of
diquarks is taken into account by diquark form factors.  These
phenomenological vertex functions multiply each $n$-point
contribution, that is, those Feynman graphs where $(n-2)$ gauge bosons
couple to the diquark.  The particular choice for space-like $Q^2$
\begin{eqnarray}
F_{S}^{(3)} (Q^2) & = & \delta_{S} \, \frac{Q_{S}^2}{Q_{S}^2 +
Q^2} , \nonumber \\
F_{V}^{(3)} (Q^2) & = & \delta_{V}
\left(\frac{Q_{V}^2}{Q_{V}^2 +
Q^2}\right)^2, \label{form3}
\end{eqnarray}
for 3-point functions and
\begin{eqnarray}
F_{S}^{(n)} (Q^2) & = & a_{S} F_{S}^{(3)} (Q^2) , \nonumber \\
F_{V}^{(n)} (Q^2) & = & a_{V} F_{V}^{(3)} (Q^2) \left(\frac{Q_{V}^2}
{Q_{V}^2 + Q^2} \right)^{(n-3)}, \label{form4}
\end{eqnarray}
for n-point functions ($n \geq 4$) ensures that in the limit $Q^2
\rightarrow \infty$ the scaling behavior of the diquark model turns
into that of the pure quark HSP. The factor $\delta_{S (V)} = \alpha_s
(Q^2) / \alpha_s (Q^2_{S (V)})$ ($\delta_{S (V)} = 1$ for $Q^2 \leq
Q^2_{S (V)}$) provides the correct powers of $\alpha_s (Q^2)$ for
asymptotically large $Q^2$.  We use the one-loop running coupling \\
$\alpha_{s} = 12 \pi / \left(25 \ln (Q^2 / \Lambda_{\sQ \sC \sD}^2
)\right)$, where $\Lambda_{\sQ \sC \sD} = 200 {\rm \hbox{ MeV}}$,
and we restrict $\alpha_{s}$ to be smaller than $0.5$.  The
$a_{D}$ are strength parameters which allow for the possibility of
diquark excitation and break-up in intermediate states where
diquarks can be far off-shell.

The parameterizations (\ref{form3}) and (\ref{form4}) are only
valid for space-like $Q^2$. For time-like arguments $s$ we have
chosen the following prescription:
\begin{eqnarray}
F_{S}^{(3)} (s) & = & \delta_{S} \, \frac{Q_{S}^2}{Q_{S}^2 -
s}  , \quad
F_{S}^{(n)}  (s) =  a_{S} F_{S}^{(3)} (s), \,\, n>3\, , \nonumber\\
F_{V}^{(3)} (s) & = & - \delta_{V} \left(\frac{Q_{V}^2}{Q_{V}^2 -
s}\right) \left(\frac{Q_{V}^2}{Q_{V}^2 + s}\right), \nonumber \\
F_{V}^{(n)} (s) & = & - a_{V} F_{V}^{(3)} (s)
\left(\frac{Q_{V}^2}{Q_{V}^2 - s}\right) \left(\frac{Q_{V}^2}
{Q_{V}^2 + s} \right)^{(n-4)}\!\!\!\!\!\! ,\,\, n>3\, .
\nonumber\\ \label{formtime}
\end{eqnarray}
This choice of diquark form factors guarantees the correct
asymptotic behavior but leads to unphysical poles.  The somewhat
more complicated form of the vector-diquark form factors has been
chosen to reduce the power of these poles as compared to the
straightforward analytical continuation $Q^2 \rightarrow -s$.  To
avoid the unphysical poles we keep the time-like diquark form
factors constant once they have reached a certain value, $c_0$,
for which we take 1.3.  However, our results are quite insensitive
to the exact value of $c_0$, since it only plays a role in the
endpoint regions, which are suppressed by the diquark wave
functions (\ref{huangv}).  We want to emphasize that the
continuation of the diquark form factors from space-like to
time-like arguments is not unique, since the underlying dynamics
is unknown. Our prescription is the same as the one adopted in
Ref.~\cite{time} for the simultaneous description of $\gamma\gamma
\rightarrow p\bar{p}$, the electromagnetic proton form factor in
the time-like domain, and the decay $\eta_{c} \rightarrow
p\bar{p}$.  The form factors in the time-like region are obviously
larger than in the space-like region, because one is closer to the
(unphysical) singularities for time-like momentum transfers,
$s>0$, than for space-like, $-Q^2=s<0$.  Experimental evidence
that this parameterization is reasonable is given for example by
the larger sizes of nucleon form factors in the time-like compared
to the space-like region, which coincide with predictions from the
quark-diquark model \cite{time}.

The complete set of parameters of the quark-diquark model is listed in
Table \ref{params}.  We emphasize that the only a priori free
parameters of the model are the constants $f_{D}, c_i$ in the
wave function (\ref{huangv}), the values of $Q_{D}^2, a_{D}$ in the
diquark form factors, and the anomalous magnetic moment of the vector
diquark $\kappa_V$.  The remaining constants $m_{q}$, $m_{D}$, and
$b^2$ which appear in the DA are fixed by the physical considerations
explained above.  The initially free parameters were fixed in
\cite{fixed} by fitting elastic electron-nucleon scattering data, and
all subsequent calculations within the model have used this set of
parameters with success.

\begin{table}
\caption{Parameters of the diquark model}
\label{params}
\begin{tabular}{ll}
\hline\noalign{\smallskip}
$m_q =  $ 330 MeV & $m_D = $ 580 MeV, for light quarks \\
& strange quarks are 150 MeV heavier \\
$b^2 = $ 0.248 $\mbox{GeV}^{-2}$ \hspace*{2mm} & \\
\noalign{\smallskip}\hline\noalign{\smallskip}
$f_{S}^{B} = $ 73.85 MeV & $f_{V}^{B} = $ 127.7 MeV \\
$Q_{S}^2 = $ 3.22 $\mbox{GeV}^2$ & $Q_{V}^2 = $ 1.50 $\mbox{GeV}^2$ \\
$a_{S} = $ 0.15 & $a_{V} = $ 0.05 \\
$c_1 = $ 0 & $c_2 = $ 0, $\quad\quad$ for scalar diquarks \\
$c_1 = $ 5.8 & $c_2 = $ - 12.5, $\quad$ for vector diquarks \\
$\kappa_{V} = $ 1.39 & \\
\noalign{\smallskip}\hline
\end{tabular}
\end{table}

\subsection{Treatment of constituent masses}
\label{sec:mass}

Above, we assumed that every baryonic constituent has a four-momentum
proportional to the four-momentum of its parent hadron
\cite{anscarm}. Therefore, every constituent of a baryon $B$ carrying
momentum fraction $x\, p_B$ acquires an effective mass $x\, m_B$,
where $m_B$ is the baryon mass. Since the momentum fractions are
weighted by the hadron DA (\ref{huangv}) in the convolution integral,
Eq. (\ref{HSP}), the quark and diquark constituents carry average
masses
\begin{eqnarray}
\left< m_{\sq}^{\mbox{\tiny av}} \right> & = & \left< x \right> m_B
\approx \frac{1}{3} m_B, \label{mquark} \\
\left< m_{D}^{\mbox{\tiny av}} \right> & = & \left< (1-x) \right>
m_B \approx \frac{2}{3} m_B. \label{mdiquark}
\end{eqnarray}
We assign the effective masses $x\, m_{B}$ to the on-shell partons
at the external legs of the Feynman diagrams for the calculation
of the hard-scattering amplitudes $\hat{T}$.  To internal lines we
assign masses according to the momentum fractions they carry,
following the same argumentation.  For a detailed explanation of
the assignment of masses to internal propagators we refer to Ref.
\cite{CW}.  The hard-scattering amplitudes are then expanded in
powers of the small parameter $(m_B/\sqrt{\sss} )$ up to
next-to-leading order, at fixed center-of-mass scattering angle
$\hat{\theta}$.  The result is reexpressed in terms of massless
Mandelstam variables, $\sss,\,\tt,$ and $\uu$, which are related
to the usual massive ones, $s,\,t,$ and $u$, by
\begin{eqnarray}
s & = & \sss \left[ 1 + \mathcal{O}  \left( (m_B/\sqrt{\sss} )^2
\right) \right],  \,\,\, \sss = 4 q^2, \\
t & = & \tt \left[ 1 + \mathcal{O} \left( (m_B/\sqrt{\sss} )^2
\right)\right],  \,\,\, \tt = - 2 q^2 \left[ 1 - \cos \hat{\theta}
\right], \nonumber \\
u & = & \uu \left[ 1 + \mathcal{O}  \left( (m_B/\sqrt{\sss} )^2
\right) \right],   \,\,\, \uu = - 2 q^2 \left[ 1 + \cos
\hat{\theta} \right], \nonumber
\end{eqnarray}
with  photon center-of-mass momentum $q$.

We emphasize that this treatment of constituent masses in the
hard-scattering amplitude does not require the introduction of new
mass parameters, contrary to the prescription used in
\cite{gamgampp,Compton}.  Our mass treatment is consistent in the
sense, that it preserves $U(1)$ gauge invariance with respect to
the photon and $SU(3)$ gauge invariance with respect to the
gluons.  As we will see below, it also provides the correct
crossing relations between the (hadronic) amplitudes for
two-photon annihilation into baryons, Eqs.  (\ref{cmsampscr}), and
those for Compton scattering off baryons.  Moreover, by including
mass corrections up to $\mathcal{O} (m_B/\sqrt{\sss} )$, not only
vector diquarks but also quarks can change their helicity.  Thus
also the quark-scalar diquark state is able to contribute to
helicity-flip amplitudes.  Such contributions have been neglected
throughout in previous work
\cite{gamgampp,Compton,fixed,time,Ja94,KSG96,KSPS97}.  The
inclusion of helicity-flip contributions from the quark-scalar
diquark system naturally leads to more pronounced polarization
effects for observables which require baryonic helicity flips.

\subsection{The elementary $\gamma \gamma \rightarrow q  D \bar{q}
\bar{D}$ amplitude}\label{sec:ampres}

There are altogether 60 Feynman graphs (30 containing S diquarks, 30
with V diquarks) which contribute to the elementary hard scattering
amplitudes $\hat{T}$ for $\gamma \gamma \rightarrow q  D \bar{q}
\bar{D}$. $\hat{T}$ has the general structure
\begin{equation}
\hat{T}_{\{\lambda\} } =  e_{\sq}^2 \hat{T}^{(3, D)}_{\{\lambda\} } +
e_{\sq} e_{D} \hat{T}^{(4,D)}_{\{\lambda\} } + e_{D}^2
\hat{T}^{(5,D)}_{\{\lambda\} },
\end{equation}
where $e_{\sq}$ and $e_{D}$ are the electrical charges of quarks and
diquarks, respectively.  Each $n$-point contribution $\hat{T}^{(n)}$
is found from a separately gauge-invariant set of Feynman diagrams and
has to be multiplied with the appropriate diquark form factors, Eq.
(\ref{formtime}):
\begin{eqnarray}
\hat{T}^{(3, D)}_{\{\lambda\} }  & = & \overline{T}^{(3, D)}_i
\left(x_1,y_1;\tt,\uu\right) F_D^{(3)}\left(x_2 y_2 \sss\right),
\nonumber \\
\hat{T}^{(4, D)}_{\{\lambda\} }  & = & \frac{\overline{T}^{(4,
D)}_i\left(x_1,y_1;\tt,\uu\right) }{\left(g_1^2 + i \varepsilon
\right) \left(g_2^2 + i \varepsilon' \right)}  F_D^{(4)}\left( (x_1
y_2 + x_2 y_1) \sss/2 \right), \nonumber \\
\hat{T}^{(5, D)}_{\{\lambda\} }  & = & \overline{T}^{(5, D)}_i
\left(x_1,y_1;\tt,\uu\right) F_D^{(5)}\left( (x_1 y_1 + x_2 y_2)
\sss\right),
\end{eqnarray}
where $i = 1,\dots, 6$ labels the helicities according to Eq.
(\ref{cmsampscr}). Above, we factored the gluon propagators $g_1^2$
and $g_2^2$ out of the four-point functions:
\begin{eqnarray}
g_1^2 & = & x_2 y_1 \,\hat{u} + x_1 y_2 \hat{t}, \nonumber \\
g_2^2 & = & x_2 y_1 \,\hat{t} + x_1 y_2 \hat{u}. \label{gluon}
\end{eqnarray}
As we will see below, in Compton scattering off baryons, which is
related to the above process by $\sss \leftrightarrow \tt $
crossing, these gluon propagators can go on shell.  In this case,
the poles arising in the convolution integral (\ref{HSP}) have to
be treated with care when performing the integration numerically.
However, in the process $\gamma \gamma \rightarrow B \bar{B}$
there are no propagator singularities and the convolution
integrations can be carried out straightforwardly.  The explicit
expressions for the amplitudes $\overline{T}^{(n, D)}_i$ are
listed in Tables \ref{scalarBB} and \ref{vectorBB} for scalar and
vector diquarks, respectively. Note that the diquark form factors
(\ref{formtime}) provide additional inverse powers of $\sss$.

\begin{table}
\caption{Elementary helicity amplitudes for the subprocess
$\gamma \gamma \rightarrow q S  \bar{q} \bar{S}$ including mass
corrections of
 $\mathcal{O} (m_{B}/\sqrt{\sss})$. For convenience we have chosen
 $\hat{t}$ and $\hat{u}$ as arguments of these amplitudes ($\sss = - \tt -
\uu$).}
\label{scalarBB}
\vspace*{-5mm}
\begin{displaymath}
C = (4 \pi)^2 C_F \alpha \alpha_s,
\end{displaymath}
where $C_F = \frac{4}{3}$ is the Casimir of the fundamental representation
of SU(3),
and $\alpha$ denotes the fine structure constant $\alpha \approx 1/137$.
\vspace*{-2mm}
\begin{eqnarray*}
\hline\noalign{\smallskip} \overline{T}_1^{(3, S)} (\tt,\uu) & = &
4 C   \frac{1}{\sss \sqrt{\uu \tt }} \left( \frac{\uu }{x_1 y_1 }
+ \frac{\tt }{x_2 y_2} \right)
 \\
\overline{T}_2^{(3, S)} (\tt,\uu) & = & 2 C    m_{B} \frac{
\sqrt{\sss}}{\uu
\tt } \frac{x_1 + y_1 }{x_1 y_1 } \\
\overline{T}_3^{(3, S)} (\tt,\uu) &  & \mbox{suppressed by }
\mathcal{O}((m_{B}/\sqrt{\sss})^2) \\
\overline{T}_4^{(3, S)} (\tt,\uu) & = & 2 C   m_{B}
\frac{1}{\sqrt{\sss}\, \sss \uu \tt} \frac{1 }{x_1 x_2 y_1 y_2 }
\left\{ \sss^2 x_2 \left[y_1 (x_1 + y_1) - 2 \right]
 \right. \\
& &  + \tt^2 \big[  (x_1 + y_1)^2 - 2 x_2 y_2 + 2 (x_1 y_1 - 1)\big]
\\
& & \left. + 2 \sss \tt
\big[ y_1(x_1 + y_1) - 2 x_2 \big] \right\}  \\
\overline{T}_5^{(3, S)} (\tt,\uu) & = & \overline{T}_1^{(3, S)}
(\uu,\tt) \\
\overline{T}_6^{(3, S)} (\tt,\uu) & = & 2 C    m_{B} \frac{
\sqrt{\sss}}
{\uu \tt } \frac{x_1 + y_1 }{x_2 y_2 } \\
\hline\noalign{\smallskip} \overline{T}_1^{(4, S)}(\tt,\uu) & = &
-4 C   \frac{1}{x_1 y_1 \sqrt{\uu \tt }}  \left[ 2 x_1 y_1 x_2 y_2
\sss^2 + \left(x_1 - y_1 \right)^2 \uu \tt  \right.
\\
& & \left. + x_1 y_1 (x_2 + y_2) \sss \tt \right]
 \\
\overline{T}_2^{(4, S)} (\tt,\uu) & = & 4 C    m_{B} \frac{g_{1}^2
g_{2}^2 \sqrt{\sss}}
{\uu \tt } \frac{x_1 + y_1 }{x_1 y_1 } \\
\overline{T}_3^{(4, S)} (\tt,\uu) &  & \mbox{suppressed by }
\mathcal{O}((m_{B}/\sqrt{\sss})^2) \\
\overline{T}_4^{(4, S)} (\tt,\uu) & = & 2 C    m_{B}
\frac{\sqrt{\sss}}{ x_1 y_1 \uu \tt} \left\{ \sss^2 x_1 y_1 x_2
\left[ (y_1 - y_2) (x_1 + y_1) - 2 \right] \right.
\\ & & + \tt^2 \big[ (x_1 + y_1)^3 - 8 x_1 y_1 \big]
 - \sss \tt \big[ - (x_1 + y_1)^3   \\
& & \left.  + x_1 y_1 (x_1^2 - y_1^2) + 2 x_1 y_1 (x_2 - y_2 + 4)
\big] \right\}
\\
\overline{T}_5^{(4, S)} (\tt,\uu) & = & \overline{T}_1^{(4, S)}
(\uu,\tt) \\
\overline{T}_6^{(4, S)} (\tt,\uu) & = & - 2 C    m_{B} \frac{
\sqrt{\sss} \, \sss^2}
{\uu \tt } (x_1 + y_1) (x_1 y_2 + y_1 x_2) \\
\hline\noalign{\smallskip}
\overline{T}_1^{(5, S)} (\tt,\uu) & = & -4 C   \frac{1}{\sqrt{ \uu
\tt }}
 \frac{1 }{x_1 y_1 } \\
\overline{T}_2^{(5, S)} (\tt,\uu) & = & 2 C    m_{B} \frac{
\sqrt{\sss}}
{\uu \tt } \frac{x_1 + y_1 }{x_1 y_1 } \\
\overline{T}_3^{(5, S)} (\tt,\uu) &  & \mbox{suppressed by }
\mathcal{O}((m_{B}/\sqrt{\sss})^2) \\
\overline{T}_4^{(5, S)} (\tt,\uu) & = & -2 C   m_{B}
\frac{1}{\sqrt{\sss} \, \sss \uu \tt} \frac{x_1 + y_1 }{x_1^2
y_1^2 } \left[ \uu \tt (x_2 + y_2)
+ \sss^2 x_1 y_1 \right] \\
\overline{T}_5^{(5, S)} (\tt,\uu) & = & \overline{T}_1^{(5, S)}
(\uu,\tt) \\
\overline{T}_6^{(5, S)} (\tt,\uu) & = & 2 C    m_{B} \frac{
\sqrt{\sss}} {\uu \tt } \frac{x_1 + y_1 }{x_1 y_1 } \\
\hline\noalign{\smallskip}
\end{eqnarray*}
\end{table}

\begin{table}
\caption{Elementary helicity amplitudes for the subprocess
$\gamma \gamma \rightarrow q V  \bar{q} \bar{V}$ including mass
corrections of
 $\mathcal{O} (m_{B}/\sqrt{\sss})$.
The 5-point functions do not contribute at all for vector
diquarks, since they are suppressed by
$\mathcal{O}((m_{B}/\sqrt{\sss})^2)$
 or higher.
}
\label{vectorBB}
\vspace*{-2mm}
\begin{displaymath}
C = (4 \pi)^2 C_F \alpha \alpha_s
\end{displaymath}
as in Table \ref{scalarBB}.
\begin{eqnarray*}
\hline\noalign{\smallskip}
\overline{T}_1^{(3, V)} (\tt,\uu) & = & -2 C
\frac{\kappa_{V}}{m_{B}^2
\sqrt{\uu \tt }} \left( \frac{\uu }{x_1 y_1 } + \frac{\tt }{x_2 y_2}
\right)
 \\
\overline{T}_2^{(3, V)} (\tt,\uu) & = & C    \frac{2 + 3
\kappa_{V}}{m_{B} }
\frac{ \sqrt{\sss} \,\sss}{\uu \tt } \frac{x_1 + y_1 }{x_1 y_1 } \\
\overline{T}_3^{(3, V)} (\tt,\uu) &  & \mbox{suppressed by }
\mathcal{O}((m_{B}/\sqrt{\sss})^2) \\
\overline{T}_4^{(3, V)} (\tt,\uu) & = & - C   \frac{1 }{m_{B}  }
\frac{1}{\sqrt{\sss}\,\uu \tt} \frac{1 }{x_1 x_2 y_1 y_2 } \bigg\{
(2 + 3 \kappa_{V}) x_2  \\
& & \times (x_1 y_1 + y_2^2) \sss^2  - 4 (x_2 - y_1)   \\
& & \times \left[ y_2 \uu \tt \frac{\scriptstyle 1}
{\scriptstyle 2}(2+3 \kappa_{V}) +
\kappa_{V} \uu (\uu - y_1 \tt) \right. \\
& & \left. -  \frac{1}{2 } (x_2 -y_2)
\tt (\tt - \kappa_{V} \uu + \frac{3}{ 2 }
\kappa_{V} \tt ) \right]   \\
& &  - \kappa_{V} (x_1 + y_1) \uu \tt +  \kappa_{V} (\tt - \uu)
(x_2 \uu - y_2 \tt)   \\
& & - 2  \kappa_{V} (x_2 \tt + y_1 \uu)(y_2 \tt + x_1 \uu)   \\
& & + 2 \kappa_{V} \uu (x_1 y_1 - x_2 y_2)(\tt - \uu) \bigg\} \\
\overline{T}_5^{(3, V)} (\tt,\uu) & = & \overline{T}_1^{(3, V)}
(\uu,\tt) \\
\overline{T}_6^{(3, V)} (\tt,\uu) & = & C    \frac{1 }{m_{B}}
\frac{ \sqrt{\sss} \, \sss}{\uu \tt } \left[ 2  (1 + \kappa_{V})
\frac{1}{x_1 y_1}
\left( \frac{x_1 y_2 }{x_2 } + \frac{y_1 x_2}{y_2 } \right) \right. \\
& & + \left. \frac{ \kappa_{V}  }{x_2 y_2 } (x_1 + y_1) \right]  \\
\hline\noalign{\smallskip}
\overline{T}_2^{(4, V)} (\tt,\uu) & = & 2 C   \frac{ \kappa_{V}
 (1 - \kappa_{V}) \sqrt{\sss}}{m_{B}^3} \frac{g_{1}^2
g_{2}^2}{x_1 x_2^2 y_1 y_2^2 }
\Big[ x_1 + y_1  \\
& & - x_1 y_1 (2 + x_2 + y_2) \Big]
\\
\overline{T}_1^{(4, V)},\,\overline{T}_3^{(4, V)}\!&\!,&
\overline{T}_5^{(4, V)}
\mbox{ suppressed by } \mathcal{O}((m_{B}/\sqrt{\sss})^2) \\
\overline{T}_4^{(4, V)},\,\overline{T}_6^{(4, V)} &  &
\mbox{suppressed by } \mathcal{O}((m_{B}/\sqrt{\sss})^3) \\
\hline\noalign{\smallskip}
\vspace*{-5mm}
\end{eqnarray*}
\end{table}

\section{Results}
\label{sec:results}
\begin{figure}
\epsfig{file=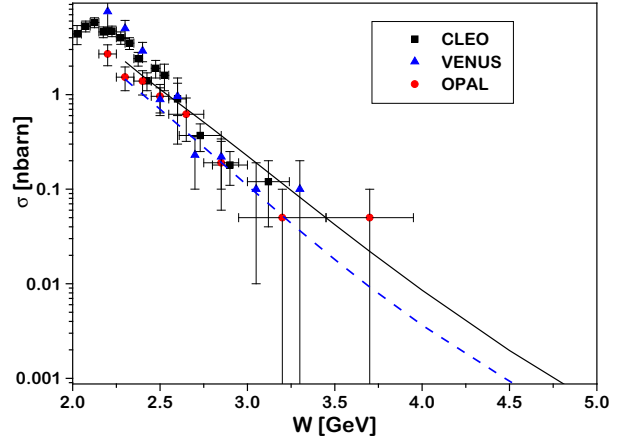,height=8.2cm,angle=270,clip=0} \caption{
Integrated cross section for $\gamma \gamma \rightarrow p \bar{p}$
($|\cos(\theta)| < 0.6$) versus $W = \sqrt{s}$. The solid line is
the complete diquark model prediction, the dashed line is the
contribution of only helicity conserving amplitudes,
$\bar{\phi}_1$ and $\bar{\phi}_5$. Data are taken from Refs.
\cite{CLEO} (CLEO), \cite{TRISTAN} (VENUS), and \cite{OPAL}
(OPAL). } \label{fig:ppbartot}
\end{figure}
\begin{figure}\mbox{ } \vspace*{3mm} \\
\epsfig{file=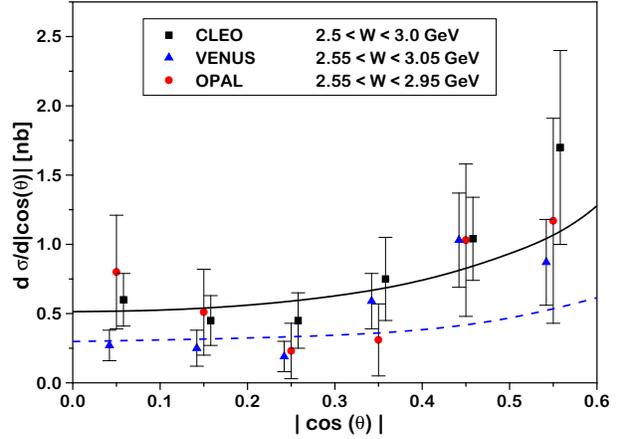,height=8.2cm,angle=270,clip=0} \caption{
Differential cross section $d \sigma(\gamma \gamma \rightarrow p
\bar{p})/d |\cos(\theta)|$) at $W = \sqrt{s} = 2.8 GeV$. Data are
taken from Refs. \cite{CLEO} (CLEO), \cite{TRISTAN} (VENUS), and
\cite{OPAL} (OPAL). Solid and dashed lines as in Fig.
\ref{fig:ppbartot}. } \label{fig:ppbarang}
\end{figure}
In the course of the past decade, large-$|t|$ cross sections for
$\gamma \gamma \rightarrow p \bar{p}$ have been measured by
various groups~\cite{CLEO,TRISTAN,OPAL}.  The diquark-model
predictions for integrated ($|\cos(\theta)| < 0.6$) as well as
differential cross sections are seen to lie well within the range
of the corresponding data (see  Figs.~\ref{fig:ppbartot} and
\ref{fig:ppbarang}, respectively).  The comparison of the solid
and the dashed lines shows that the contributions of the mass
corrections which enter via the hadronic helicity-flip amplitudes
$\bar{\phi}_{2}$, $\bar{\phi}_{4}$, and $\bar{\phi}_{6}$ are of
the same order of magnitude as that of the leading-order hadronic
helicity conserving amplitudes $\bar{\phi}_{1}$ and
$\bar{\phi}_{5}$~\footnote{Within our model $\bar{\phi}_{3}$ is
suppressed by $\mathcal{O} (m_B^2/\sss)$}.  The dashed line
corresponds also approximately to the results given in
Ref.~\cite{time}, in which quark masses have been neglected
completely and the vector-diquark mass has been taken as a fixed
parameter.  In contrast to our results, the perturbative
predictions of the pure quark HSP~\cite{FMN85,FZOZ89} are at least
one order of magnitude below the data.

\begin{figure}
\epsfig{file=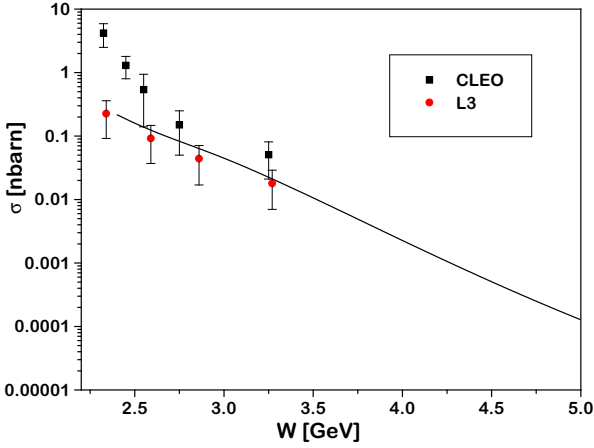,height=8.2cm,angle=270,clip=0}
\caption{ Integrated cross section for $\gamma \gamma \rightarrow
\Lambda \bar{\Lambda}$ ($|\cos(\theta)| < 0.6$) versus $W =
\sqrt{s}$. Data are taken from Refs. \cite{CLEO2} (CLEO) and
\cite{L3} (L3). }
\label{fig:llbartot}
\end{figure}
\begin{figure}
\epsfig{file=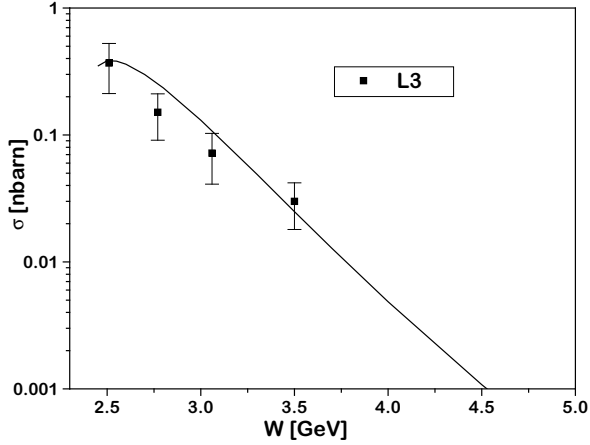,height=8.2cm,angle=270,clip=0} \caption{
Integrated cross section for $\gamma \gamma \rightarrow \Sigma^0
\bar{\Sigma}^0$ ($|\cos(\theta)| < 0.6$) versus $W = \sqrt{s}$.
Data are taken from Ref. \cite{L3} (L3). } \label{fig:s0s0bartot}
\end{figure}

\begin{figure}
\epsfig{file=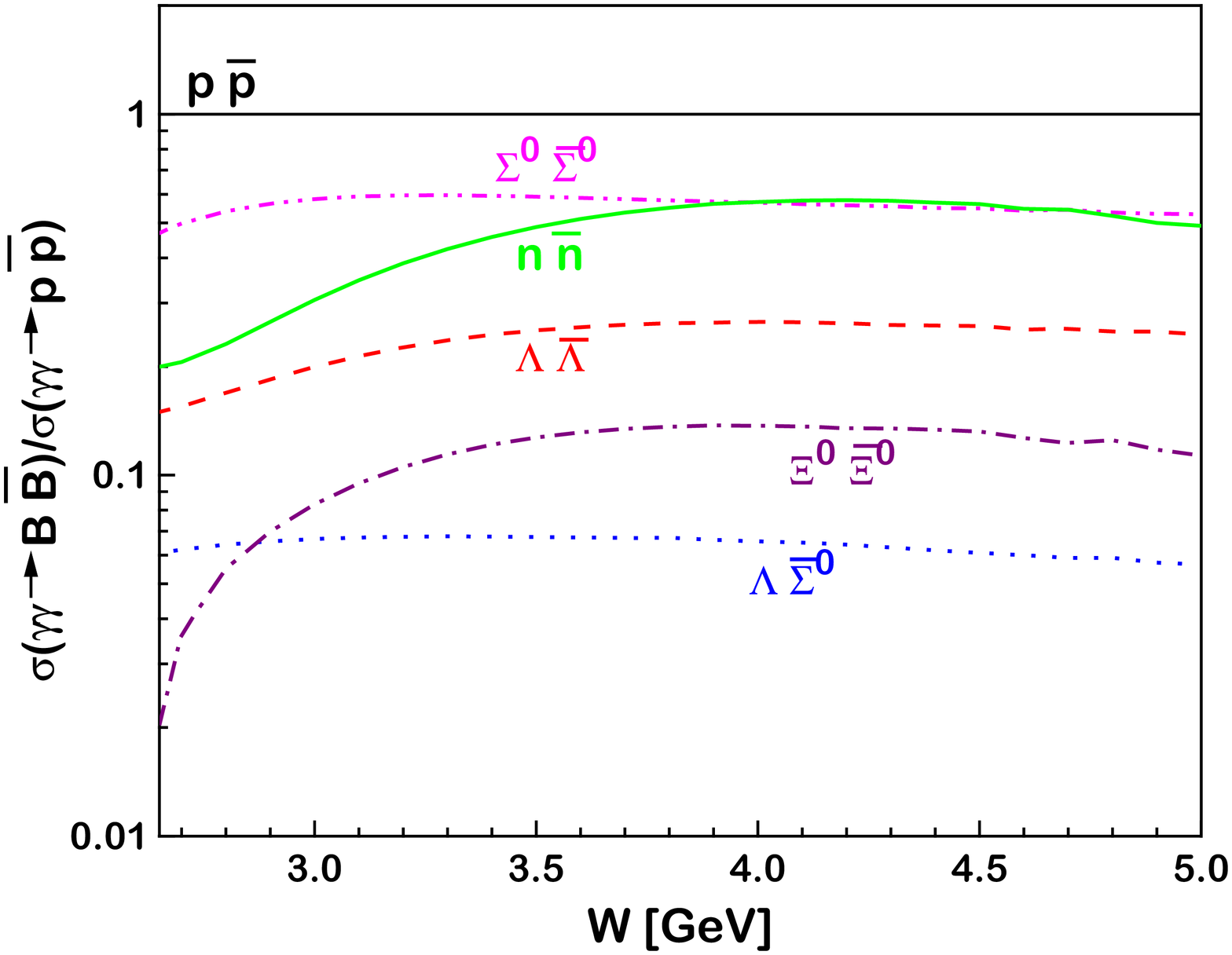,height=8.2cm,angle=270,clip=0}
\caption{Diquark model predictions for $\gamma \gamma$-annihilation
into baryon-antibaryon pairs. Shown are ratios of integrated cross
sections $\sigma(\gamma \gamma \rightarrow B \bar{B})/\sigma(\gamma
\gamma \rightarrow p \bar{p})$ ($|\cos(\theta)| < 0.6$) versus $W =
\sqrt{s}$ for neutral baryons. }
\label{fig:allrat0}
\end{figure}
\begin{figure}
\epsfig{file=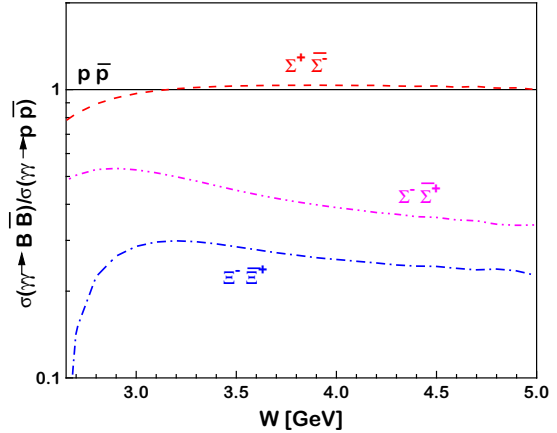,height=8.2cm,angle=270,clip=0}
\caption{Diquark model predictions for $\gamma \gamma$-annihilation
into baryon-antibaryon pairs. Shown are ratios of integrated cross
sections $\sigma(\gamma \gamma \rightarrow B \bar{B})/\sigma(\gamma
\gamma \rightarrow p \bar{p})$ ($|\cos(\theta)| < 0.6$) versus $W =
\sqrt{s}$ for charged baryons. }
\label{fig:allratpm}
\end{figure}
With the SU(3)-symmetric flavor parts of the quark-diquark wave
functions given in Tab.~\ref{flav} and the flavor-dependent
distribution amplitude, Eq.~(\ref{huangv}), we are in the position
to treat not only $p\bar{p}$ production, but also the production
of other octet-baryon pairs without introducing new parameters.
For the $\Lambda$ and $\Sigma^0$ channels integrated cross-section
data have been published very recently~\cite{CLEO2,L3}.  As shown
in Fig.~\ref{fig:llbartot} our predictions for the $\Lambda$
channel agree favorably with the most recent L3 measurements which
are somewhat below the older CLEO data points.  Comparable good
agreement with the L3 data is also achieved for the $\Sigma^0$
channel (cf. Fig.~\ref{fig:s0s0bartot}).  For the other
octet-baryon channels we have no experimental information as yet,
apart from a statement by the L3 collaboration that the
$\Lambda\bar{\Sigma}^0+\Sigma^0\bar{\Lambda}$ cross section is
below their present detection accuracy.  In order to have a
guideline we have therefore plotted the ratios of integrated
$B\bar{B}$ cross sections ($|\cos(\theta)| < 0.6$) to the
$p\bar{p}$ cross section.  These ratios are shown in
Fig.~\ref{fig:allrat0} and Fig.~\ref{fig:allratpm} for neutral and
charged baryons, respectively.  From these figures one observes
that it may be feasible to measure also other hyperon channels
like $\Sigma^+\bar{\Sigma}^-$, $\Sigma^-\bar{\Sigma}^+$, or
$\Xi^-\bar{\Xi}^+$, since the corresponding cross sections are
predicted to be of the same size or even larger than the
$\Lambda\bar{\Lambda}$ cross section. Such data could be very
useful to determine the amount of SU(3) flavor symmetry breaking
in the baryon distribution amplitudes. At asymptotically large
momentum transfers $|t|,\; |u| \rightarrow \infty$, where the
diquarks dissolve into quarks the three-quark distribution
amplitudes of the octet baryons satisfy exact SU(6) spin-flavor
symmetry. This leads to relations between the various production
cross sections for flavor octet and decuplet
baryons~\cite{farrar}.  SU(3) flavor symmetry, in particular
$U$-spin invariance (which is the symmetry under interchange of
$d$ and $s$ quarks) alone, implies already that
\begin{eqnarray}
\sigma(\gamma \gamma \rightarrow p  \bar{p})\;\;\;\,\; & = &
\sigma(\gamma \gamma
\rightarrow \Sigma^+
 \bar{\Sigma}^-), \\
\sigma(\gamma  \gamma \rightarrow n  \bar{n})\;\;\;\, \;& = &
\sigma(\gamma  \gamma \rightarrow \Xi^0
 \bar{\Xi}^0), \\
\sigma(\gamma  \gamma \rightarrow \Sigma^-\!
 \bar{\Sigma}^+) & = & \sigma(\gamma  \gamma \rightarrow \Xi^-
\bar{\Xi}^+).
\end{eqnarray}
Two more relations follow from U-spin invariance.  These, however,
assume only a simple form on the amplitude level:
\begin{eqnarray}
\mathcal{M}^{\Lambda\bar{\Sigma}^0} &=& \sqrt{3}
(\mathcal{M}^{\Lambda\bar{\Lambda}}-
\mathcal{M}^{n\bar{n}}), \\
\mathcal{M}^{\Sigma^0\bar{\Sigma}^0} &=&
3 \mathcal{M}^{\Lambda\bar{\Lambda}}-
2 \mathcal{M}^{n\bar{n}}.
\end{eqnarray}
These relations hold for each helicity amplitude.  At finite
momentum transfers deviations from SU(3)-flavor symmetry have
their origin in the different baryon masses and the flavor
dependence of the baryon DAs.  This is precisely the observation
which can be made from Figs.~\ref{fig:allrat0} and
\ref{fig:allratpm}.  The larger the mass difference of the
produced baryon-antibaryon pair the bigger the violation of SU(3)
flavor symmetry.  The asymptotic SU(6) spin-flavor symmetry gives
rise to additional relations between the amplitudes for octet-
and decuplet-baryon production~\cite{farrar}.  A systematic
breaking of SU(6) spin-flavor symmetry down to SU(3) flavor
symmetry is inherent in the diquark model and is caused by the
assumption of flavor dependent scalar and vector-diquark masses,
different vertex form factors, and different values for $f_{S}$
and $f_{V}$. This kind of symmetry breaking could be investigated
by comparing the production of octet and decuplet baryons.

Very recently, the production of baryon pairs has also been
investigated within the generalized parton picture
in~\cite{dikrovo}. The authors of Ref.~\cite{dikrovo} analyze the
handbag contribution to $\gamma\gamma \rightarrow B\bar{B}$ and
obtain an expression for the differential $\gamma\gamma
\rightarrow B\bar{B}$ cross section which, after some simplifying
assumptions, contains only one effective $s$-de\-pen\-dent form
factor.  After fixing this form factor by means of integrated
$p\bar{p}$ cross-section data, the authors employ isospin and
$U$-spin invariance to give amplitude ratios
$r_{B}=\mathcal{M}^{B\bar{B}}/\mathcal{M}^{p\bar{p}}$ for the
other octet baryons $B$.  In this way they obtain similar good
agreement with the integrated $\Lambda\bar{\Lambda}$ and
$\Sigma^0\bar{\Sigma}^0$ cross section data as we do.  In a
similar spirit the time-reversed process $p\bar{p}\rightarrow
\gamma\gamma$ has been considered in Ref.~\cite{FRSW02}.  In this
paper, however, it has been attempted to model directly the
time-like double distributions which describe the transition of
the $p\bar{p}$ to the $q\bar{q}$ pair.  For $s=10$~GeV$^2$ the
authors predict an integrated cross section ($|\cos(\theta)| <
0.7$) of $0.25 \times 10^{-9}$\,fm$^2$. We find a considerably
larger cross section, namely $0.14 \times 10^{-7}$~fm$^2$, which,
however, agrees with the estimate of Diehl et
al.~\cite{dikrovo,kroprivat}.  Such predictions may be of interest
regarding the proposal to build an antiproton storage ring at GSI.

\section{The crossed process $\gamma B \rightarrow \gamma B$}
\label{sec:compton}

Although the focus of the present work is baryon pair production
in two-photon collisions, we want to briefly comment on the
crossed process, Compton scattering off baryons.  Predictions for
real and virtual Compton scattering off protons have already been
given in Ref.~\cite{KSG96} for the current parameterization of the
diquark model.  In the following we will briefly summarize how our
improved treatment of constituent masses affects the results for
real Compton scattering.

Following the common convention \cite{rosti76} we denote the six
independent helicity amplitudes for Compton scattering
$\mathcal{M}_{\lambda_{2},\,\lambda_f;\,\lambda_1,\,\lambda_i}$
($\lambda_j,\,j = 1,2 \dots$ helicities of the incoming and outgoing
photons, respectively; $\lambda_{i},\,\lambda_{f} \dots$ helicities
of the incoming and outgoing baryon, respectively) by
\begin{eqnarray}
\phi_1 & = & \mathcal{M}_{1,\,\frac{1}{2};\,1,\,\frac{1}{2}},
\nonumber  \\
\phi_2 & = & \mathcal{M}_{-1,\,-\frac{1}{2};\,1,\,\frac{1}{2}},
\nonumber \\
\phi_3 & = & \mathcal{M}_{-1,\,\frac{1}{2};\,1,\,\frac{1}{2}},
\nonumber \\
\phi_4 & = & \mathcal{M}_{1,\,-\frac{1}{2};\,1,\,\frac{1}{2}},
\nonumber  \\
\phi_5 & = & \mathcal{M}_{1,\,-\frac{1}{2};\,1,\,-\frac{1}{2}},
\nonumber \\
\phi_6 & = &
\mathcal{M}_{-1,\,\frac{1}{2};\,1,\,-\frac{1}{2}}.\label{cmsamps}
\end{eqnarray}
These amplitudes are related to those for the crossed reaction $\gamma
\gamma \rightarrow B \bar{B}$ (see Eq.  (\ref{cmsampscr})) via
the crossing relations:
\begin{eqnarray}
\label{cross}
\overline{\phi}_1 & = & \phi_1 + 2 m_{B}
\sqrt{\frac{\hat{u}}{\hat{s}\,\hat{t}}}\, \phi_4,  \nonumber \\
\overline{\phi}_2 & = & \phi_2 - 2 m_{B}
\sqrt{\frac{\hat{u}}{\hat{s}\,\hat{t}}}\, \phi_3,  \nonumber \\
\overline{\phi}_3 & = & \phi_3 + m_{B}
\sqrt{\frac{\hat{u}}{\hat{s}\,\hat{t}}} \,\left( \phi_2 - \phi_6
\right), \nonumber \\
\overline{\phi}_4 & = & \phi_4 - m_{B}
\sqrt{\frac{\hat{u}}{\hat{s}\,\hat{t}}} \,\left( \phi_1 + \phi_5
\right), \label{crossendrel} \\
\overline{\phi}_5 & = & -\phi_5 - 2 m_{B}
\sqrt{\frac{\hat{u}}{\hat{s}\,\hat{t}}}\, \phi_4,  \nonumber \\
\overline{\phi}_6 & = & -\phi_6 - 2 m_{B}
\sqrt{\frac{\hat{u}}{\hat{s}\,\hat{t}}}\, \phi_3, \nonumber
\end{eqnarray}
where $\overline{\phi}_i = \overline{\phi}_i \left(\tt,\uu
\right)$ and $\phi_i = \phi_i \left(\sss,\uu\right)$.  That is, we
exchange $\sss \leftrightarrow \tt$ (if $\sss$ occurs as argument
of a square root it has to be replaced by $|\tt|$).  The above
relations are expanded up to first order in the baryon mass.

We have used these crossing relations to check our analytical
expressions for the elementary $\gamma \gamma \rightarrow q D
\bar{q} \bar{D}$ and $\gamma q D \rightarrow \gamma q D$
amplitudes which have been calculated separately.  Since the
explicit calculations agree with the crossing relations
(\ref{cross}), we refrain from listing the various amplitudes
contributing to $\gamma B \rightarrow \gamma B$. They can be read
off from the expressions for $\gamma \gamma \rightarrow B \bar{B}$
listed in Tables \ref{scalarBB} and \ref{vectorBB}, with the
appropriate replacements $\sss \leftrightarrow \tt$.  A further
check of our analytical expressions is the comparison in the
massless limit with the earlier results of
Refs.~\cite{Compton,KSG96}.

The numerical integrations in the convolution integral
(\ref{HSP}) deserve special care, since the gluon-propagators
appearing in the 4-point functions,
\begin{eqnarray}
g_1^{-2} & = & (x_2 y_1 \,\hat{u} + x_1 y_2 \hat{s})^{-1}, \nonumber \\
g_2^{-2} & = & (x_2 y_1 \,\hat{s} + x_1 y_2 \hat{u})^{-1}, \label{gluonC}
\end{eqnarray}
can go on-shell. The resulting propagator poles have to be treated
with a principal value prescription,
\begin{equation}
\frac{ 1}{g^2 + i \varepsilon } = \wp \left( \frac{1 }{g^2 } \right)
- i \pi \delta(g^2).
\end{equation}
Here we do not go into the technical details of implementing this
prescription numerically, but rather refer the reader to \cite{CW} for
further explanation.  We only want to emphasize, that our numerical
results are absolutely stable for both real and imaginary parts.
\begin{figure}
\epsfig{file=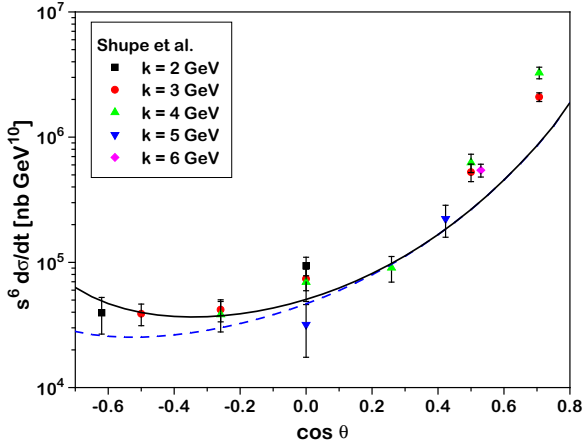,height=8.2cm,angle=270,clip=0}
\caption{Scaled differential cross section $s^6\ d \sigma(\gamma p
\rightarrow \gamma p)/d t$.  Diquark-model predictions are given
for the photon lab energy $k=4$~GeV ($W = \sqrt{s} = 2.9$~GeV)
with all (solid line) and only hadronic helicity conserving
amplitudes (dashed line) taken into account.  Data for various
photon lab energies are taken from Ref.~\cite{shu79}}
\label{fig:comptonang}
\end{figure}

As can be seen from Fig.~\ref{fig:comptonang} the predictions of
the diquark model agree nicely with the few available differential
cross-section data at intermediately large momentum transfers.
According to a recent study~\cite{brooksdix} it seems to be
unlikely, on the other hand, that the pure quark HSP is able to
account for these data.  Good results, however, are also obtained
within the generalized parton picture with its dominant
contribution being given by the handbag
diagram~\cite{rad98,kroll2}.  In this case only one parton
undergoes a hard scattering.  The (soft) emission and reabsorption
of the parton by the hadron is described by generalized parton
distributions which may be modelled by the overlaps of light-cone
wave functions, since the change of the parton momentum is
space-like.  It is not surprising that our diquark model which is
based on the hard-scattering picture yields results similar to the
soft overlap mechanism.  It indicates that part of the
multidimensional overlap integral is absorbed into diquark form
factors which, in our model, are parameterized phenomenologically
over the whole momentum-transfer range.

There are, nevertheless, qualitative differences between the
diquark model and the overlap mechanism which appear, in
particular, in polarization observables.  As a representative
example let us, e.g., mention the initial state helicity
correlation $A_{LL}$.  Whereas the soft overlap yields positive
values for this spin observable~\cite{kroll2}, we obtain a
negative result within the diquark model.  This difference can be
understood easily.  The comparison of the solid and the dashed
lines in Fig.~\ref{fig:comptonang} reveals that the (leading
order) hadronic helicity conserving amplitudes $\phi_{1}$ and
$\phi_{5}$ are dominant.  Mass corrections, which enter via the
hadron-helicity-flip amplitudes $\phi_{2}$, $\phi_{4}$, and
$\phi_{6}$, only start to play a role in the backward region. Thus
$A_{LL}$ is roughly given by the difference
$(|\phi_{1}|^2-|\phi_{5}^2|)$.  Considering the analytical
expressions for these amplitudes (cf. Tabs.~\ref{scalarBB} and
\ref{vectorBB} with $\sss \leftrightarrow \tt$ interchanged),
specifically the 3-point contributions, which are actually the
most important ones, we see that $(|\phi_{1}|^2-|\phi_{5}|^2)
\propto (1/(x_{2}^2 y_{2}^2)-1/(x_{1}^2 y_{1}^2))
(\sss^2-\uu^2)/(\tt^2 \sss |\uu|)$. For our choice of
quark-diquark distribution amplitudes the average values of the
momentum fractions are $<x_{1}>=<y_{1}>=1-<x_{2}>=1-<x_{2}>\approx
1/3$ so that one can already conclude that $A_{LL}$ should be
negative. This is indeed the result of our numerical calculations.
These considerations illustrate also why the soft overlap
mechanism provides a positive $A_{LL}$. Within such an approach
the active quark carries nearly all of the baryon momentum
($x_{1},y_{1}\approx 1$) and hence $(1/(x_{2}^2
y_{2}^2)-1/(x_{1}^2 y_{1}^2))$ becomes positive.  By similar
qualitative arguments we expect the spin-transfer parameter
$D_{LL}$ to be close to $1$ and the photon asymmetry $\Sigma$ to
be close to $0$.  Our numerical calculations confirm these
estimates.  They show in addition that polarization observables
are, unlike spin averaged quantities, very sensitive to the choice
of the quark-diquark distribution amplitudes. Experimental data
for such observables could thus be helpful to determine the
(phenomenological) quark-diquark distribution amplitudes
unambiguously.

\section{Final remarks}
\label{sec:final}

In this work we have investigated exclusive two-photon reactions
at moderately large momentum transfer within a quark-diquark
model. This approach is a modification of the pure quark
hard-scattering picture where baryons are treated as systems of
quarks and diquarks.  The work continues and extends the
systematic study of photon-induced exclusive hadronic reactions
\cite{gamgampp,Compton,fixed,time,KSG96,KSPS97,fizb,CW,mass}
per\-formed within the same approach.  The introduction of diquarks
allows us to extend the range of validity of the pure quark HSP
down to the kinematic region of a few GeV of momentum transfer,
as this modelling accounts for nonperturbative effects present in
this region.  This is also the kinematic range which is
accessible with present-day experimental facilities.

We have extended previous calculations of the two-photon reactions
$\gamma \gamma \rightarrow B \bar{B}$ and $\gamma B \rightarrow \gamma
B$ \cite{gamgampp,Compton,time}.  Refs.~\cite{Compton} and \cite{time}
have only dealt with the proton channel, Ref.~\cite{gamgampp} has
investigated other baryon channels as well, but under the simplifying
assumption that vector diquarks can be neglected compared to scalar
ones.  In Refs.  \cite{gamgampp,Compton,time} mass effects have been
treated in a simplified way, so that the crossing relations between
the $\gamma \gamma \rightarrow B \bar{B}$ and the $\gamma B
\rightarrow \gamma B$ channels were only exactly fulfilled for the
dominant helicity-nonflip amplitudes.  In this work we have studied
all octet baryon channels with an improved treatment of mass effects
by systematically expanding in the small parameter (mass/photon
energy).  We have performed our calculations within the complete
diquark model, that is, with scalar and vector diquarks.  For the
two-photon annihilation mass corrections are found to be sizable.
Their inclusion considerably improves the agreement of the diquark
model predictions with the recent LEP data for two-photon annihilation
into $\Lambda\bar{\Lambda}$ and $\Sigma^0\bar{\Sigma}^0$.  On the
other hand, mass corrections seem to be of minor importance in Compton
scattering.  It would certainly be desirable to obtain experimental
baryon pair-production data for hyperons different from $\Lambda$ or
$\Sigma^0$.  For $\Sigma^+$, $\Sigma^-$, and $\Xi^-$ we find cross
sections of comparable size so that one may hope that corresponding
measurements could be feasible.  Such data could provide information
on the flavor dependence of baryon distribution amplitudes, and they
could help to decide whether the particular scheme of $SU(6)$
spin-flavor-symmetry breaking inherent to the diquark model is
appropriate.

Finally we want to emphasize that to the best of our knowledge the
diquark model is, aside from the generalized parton picture, the only
constituent-scattering model which is able to account for the $\gamma
\gamma \rightarrow p \bar{p}$, $\gamma \gamma \rightarrow \Lambda
\bar{\Lambda}$, $\gamma \gamma \rightarrow \Sigma \bar{\Sigma}$, and
$\gamma p \rightarrow \gamma p$ data at intermediately large momentum
transfer.  It is even more remarkable that this is achieved with the
set of model parameters that provides also a reasonable description of
other exclusive quantities, like electromagnetic nucleon
factors~\cite{fixed,time} or photoproduction cross
sections~\cite{KSPS97,CW}.  Therefore further applications of this
effective approach and studies of its underlying mechanisms are
certainly worthwhile.

\begin{appendix}

\section{Feynman Rules}

\subsection{Three-point vertices}

\begin{itemize}
\item SgS-vertex:
\begin{equation}
\I \g_{\sms} T^a_{i j} (p_1 + p_2)_{\mu}
\end{equation}
\item S$\gamma$S-vertex:
\begin{equation}
-\I \e_0 \e_{S} (p_1 + p_2)_{\mu}
\end{equation}
\item VgV-vertex:
\begin{eqnarray}
-\I \g_{\sms} T^a_{i j}&\Big\{&\g_{\alpha \beta} (p_1 + p_2)_{\mu}
\nonumber \\
&-&\g_{\mu \alpha} \left[ (1 + \kappa_{V}) p_1 - \kappa_{V} p_2
\right]_\beta \nonumber \\
&-&\g_{\mu \beta} \left[ (1 + \kappa_{V}) p_2 - \kappa_{V} p_1
\right]_\alpha \Big\}
\end{eqnarray}
\item V$\gamma$V-vertex:
\begin{eqnarray}
\I \e_0 \e_{V}&\Big\{&\g_{\alpha \beta} (p_1 + p_2)_{\mu} \nonumber \\
&-&\g_{\mu \alpha} \left[ (1 + \kappa_{V}) p_1 - \kappa_{V} p_2
\right]_\beta \nonumber \\
&-&\g_{\mu \beta} \left[ (1 + \kappa_{V}) p_2 - \kappa_{V} p_1
\right]_\alpha \Big\}
\end{eqnarray}
\end{itemize}

\subsection{Four-point vertices}

\begin{itemize}
\item $\gamma$SgS-vertex:
\begin{equation}
- 2 \I \e_0 \e_{S} \g_{\sms} T^a_{ij} \g_{\mu \nu}
\end{equation}
\item $\gamma$S$\gamma$S-vertex:
\begin{equation}
2 \I \e^2_0 \e^2_{S} \g_{\mu \nu}
\end{equation}
\item gSgS-vertex:
\begin{equation}
\I \g^2_{\sms} \left\{ T^a,T^b \right\}_{i j} \g_{\mu \nu}
\end{equation}
\item $\gamma$VgV-vertex:
\begin{equation}
\I \e_0 \e_{V} \g_{\sms} T^a_{ij} (2 \g_{\mu \nu} \g_{\alpha \beta} -
\g_{\mu \beta} \g_{\alpha \nu} - \g_{\mu \alpha} \g_{\beta \nu})
\end{equation}
\item $\gamma$V$\gamma$V-vertex:
\begin{equation}
- \I \e_0^2 \e_{V}^2 (2 \g_{\mu \nu} \g_{\alpha \beta} - \g_{\mu
\beta}
\g_{\alpha \nu} - \g_{\mu \alpha} \g_{\beta \nu})
\end{equation}
\item gVgV-vertex:
\begin{eqnarray}
- \I \g^2_{\sms} &\Big\{&\left\{ T^a,T^b \right\}_{i j} \g_{\mu \nu}
\g_{\alpha \beta}  \\
&-&\left[(1 + \kappa_{V})(T^a T^b)_{i j} - \kappa_{V}(T^b T^a)_{i j}
\right] \g_{\mu \beta} \g_{\alpha \nu} \nonumber \\
&-&\left[(1 + \kappa_{V})(T^b T^a)_{i j} - \kappa_{V}(T^a T^b)_{i j}
\right] \g_{\mu \alpha} \g_{\beta \nu} \Big\} \nonumber
\end{eqnarray}
\end{itemize}
Here
$e_{\sq}, e_{S}, e_{V}$ are the charges of quarks, scalar and vector
diquarks respectively, $e_0 = \sqrt{4 \pi \alpha}$ denotes the
elementary
charge with the fine structure constant $\alpha \approx 1/137$,
$g_{S} =
\sqrt{4 \pi \alpha_s}$ is the strong coupling constant of QCD, and
$T^a$ represent the Gell-Mann color matrices.

\end{appendix}

\section*{Acknowledgements}

C. F. Berger thanks the Paul-Urban-Stipendienstiftung for
supporting a visit at the Institute for Theoretical Phys\-ics,
University of Graz, during which part of this work has been
completed. C. F. Berger was supported in part by the National
Science Foundation, grant PHY-0098527.

\end{document}